\documentclass[aps,amsfonts,amsmath,prd,preprint,nofootinbib]{revtex4}
\usepackage{epsf,mathrsfs}
\usepackage{graphicx} % To insert figures

\newcommand{\beq}{\begin{equation}}
\newcommand{\eeq}{\end{equation}}

\begin{document}

\title{Minkowski vacua can be metastable}

\author{Jaume Garriga$^{1}$, Benjamin Shlaer$^2$, Alexander Vilenkin$^2$.}

\address{$^1$ Departament de F{\'\i}sica Fonamental i \\Institut de Ci{\`e}ncies del Cosmos, 
Universitat de Barcelona,\\
Mart{\'\i}\ i Franqu{\`e}s 1, 08028 Barcelona, Spain
\\
$^2$ Institute of Cosmology, Department of Physics and Astronomy,\\
Tufts University, Medford, MA 02155, USA}

\begin{abstract}

We investigate the recent suggestion that a Minkowski vacuum is
either absolutely stable, or it has a divergent decay rate and thus fails
to have a locally Minkowski description.  The divergence comes from boost integration over momenta of
the vacuum bubbles.  We point out that a prototypical example of
false-vacuum decay is pair production in a uniform electric field, so if
the argument leading to the divergence is correct, it should apply to this
case as well. We provide evidence that no catastrophic vacuum instability
occurs in a constant electric field, indicating that the argument cannot be
right.  Instead, we argue that the boost integration that leads to the divergence is 
unnecessary: when all possible fluctuations
of the vacuum bubble are included, the quantum state of the bubble is
invariant under Lorentz boosts.

\end{abstract}

\maketitle

\section{Introduction}

In an interesting recent paper \cite{Gia}, Dvali suggested a simple argument indicating that a Minkowski vacuum cannot be metastable.   The argument can be summarized as follows.

Suppose we have a theory admitting a zero-energy metastable (false) Minkowski vacuum which can tunnel to a negative-energy, anti-de Sitter (AdS) true vacuum through bubble nucleation.   
The tunneling process conserves energy, so the critical Coleman-De
Luccia (CDL) bubble has zero energy, with the negative energy in the
bubble interior compensated by the positive energy of the bubble wall
\cite{CdL}.  A bubble whose size is larger than critical and whose
boundary is at rest will have negative total energy. This state of
negative energy may then be created from the false vacuum together
with a compensating blob of  ordinary positive-energy particles,
perhaps with some separation between the positive and the negative
energy blobs.  

One might think that the negative-energy blob is like a ghost,
that is, a negative-energy point-particle.   
If the energy-momentum vector of the bubble is $p^\mu$, the positive
energy blob must have energy-momentum vector $-p^\mu$, and the Lorentz
invariance of the Minkowski vacuum implies that the rate for this
process per unit spacetime volume can depend only on $p^2$.  Given a
positive plus negative blob at nucleation, we can obtain other
possible configurations by applying arbitrary Lorentz boosts.  This
will give configurations with arbitrarily high positive and negative
energies, adding up to zero.  Since $p^2$ does not change under
boosts, the nucleation rate for all these configurations should be the
same.  Then the total vacuum 
decay rate, obtained by integrating over the boosts, is infinite.  The
conclusion is that the concept of a metastable Minkowski vacuum is
inconsistent; such a vacuum simply cannot exist.  We shall refer to this as the ``boost integration argument."

 As noted in \cite{Gia}, a possible caveat is that a metastable
  Minkowski vacuum can exist only for a finite time and thus cannot be
  Poincar\'e-invariant.  This 
results in a cutoff for the divergent boost integral.  (This issue
will be discussed in more detail in subsequent publications
\cite{Giga}.)  However, according to the standard analysis, the bubble
nucleation rate can be arbitrarily low, resulting in an arbitrarily
high boost cutoff and an arbitrarily large enhancement of the vacuum
decay rate.  This would still call for a drastic revision of the
standard approach. In what follows we shall use the term ``metastable Minkowski vacuum" with
the understanding that this vacuum is only locally Poincar\'e-invariant.

  It is important to note that for theories without gravity the
  argument in \cite{Gia} does not rely in any way on the existence of
  negative energies.  Given a theory with a metastable zero-energy
  vacuum, decaying to a true negative-energy vacuum, we can always add
  a constant to the Lagrangian of the theory, so that both vacua become
  positive-energy.  This of course does not change the dynamics, and
  the argument should still go through, except now the energy-momentum
  vector of the bubble $p^\mu$ should be understood as the difference
  between the energy momentum vectors of the false vacuum with and
  without a bubble.  If the phase space integral in the decay rate was
  divergent in the original theory, it will still be divergent.
  Hence, if the argument is correct, then all non-gravitational
  theories with metastable vacua should be inconsistent.

The boost integration argument is reminiscent of the problem encountered in the
early treatment of vacuum decay \cite{VKO} by Voloshin, Kobzarev, and
Okun (VKO).  If we think of a CDL bubble as materializing at rest in
some frame of reference, this nucleation event is not boost-invariant.
VKO argued that the nucleation rate should therefore include an
integration over boosts, resulting in a divergent total rate.  (They
suggested a cutoff related to the radius of the universe.)  The key
insight by Coleman \cite{Coleman} was that the bubble evolution after nucleation
 is Lorentz invariant.  Hence,
integration over boosts is unnecessary, since it amounts to multiple
counting of the same final states. 

Coleman's result is supported by an independent calculation of the
vacuum decay rate in terms of the imaginary part of the vacuum energy
density \cite{CallanColeman}.  In this approach, the issue of boost
integration does not arise.  Another piece of evidence comes from the
calculations of pair creation rate in a constant electric field
\cite{Sauter} -- a process closely analogous to vacuum decay.
Starting with the work of Heisenberg and Euler \cite{HE} and Schwinger
\cite{Schwinger}, pair creation has been studied by a variety of
methods.  Results obtained using different methods are in full
agreement and do not include any divergent boost integrals. 

Here we are going to argue that Dvali's conundrum can be resolved in a
similar way.  An expanding bubble can emit particles, so the
asymptotic states at future infinity do include positive-energy
particles with a compensating negative-energy bubble.  In fact, the
total number of emitted particles and their total energy are going to
be infinite in this limit.  Suppose the bubble were formed together
with a photon, as in the boost integration argument.  This process cannot be
distinguished from bubble formation with a subsequent emission of the
photon at about the same time.  Now, if we apply a Lorentz boost, the
emission point of the photon will move to the future (or to the past).
Large Lorentz boosts will move it to the very distant future.
Integration over boosts would thus account for all possible photon
emission points.  Now, the probability that the bubble will emit a
single photon in its entire history is zero.  As we said, the total
number of emitted photons is infinite, and all final states having
nonzero probability correspond to a certain average number of photons
emitted per unit area of the bubble wall per unit proper time.  Such
states remain unchanged under Lorentz boosts. 

Another way of looking at this is to note that the bubble wall has the
geometry of a (2+1)-dimensional de Sitter space.  Any quantum field
interacting with the field(s) of the bubble will be in a de Sitter
invariant quantum state around the wall \cite{Tanmay1}.  This quantum
state accounts for all particles that will ever be emitted by the
wall.  Since the de Sitter space is characterized by an intrinsic
temperature, things more substantial than particles will sometimes also
pop out.   There will be blobs of matter, occasional Boltzmann brains,
etc.  This will be happening all along the domain wall in a de Sitter
invariant fashion.  Now, Lorentz boosts in the bulk spacetime
correspond to de Sitter transformations on the worldvolume, and since
the quantum state is de Sitter invariant, it does not change under
boosts.   Hence, including configurations of negative and positive energies
does not break the Lorentz invariance of the final state, but is necessarily
part of the quantum description of a Lorentz invariant bubble.

%The total vacuum decay rate should include a sum over all such states.  Consider, for example, the states where the bubble is accompanied by a single photon.  We should integrate over all energies of the photon and over all spacetime locations of its emission.  Should we also integrate over Lorentz boosts?  No, since a boost would simply change the energy of the photon and its emission point.  So including the boosted configurations would amount to overcounting.

To substantiate this view, we shall investigate how the boost integration argument plays out in the case of pair production in electric field.  We shall first focus on the (1+1)-dimensional case, which is a prototypical example of vacuum decay.  Starting with a brief overview in the next section, we shall 
analyze some simple massless (1+1)D models in Section \ref{class}.  Massive particle production in (3+1)D is discussed in Section \ref{sec:massive}, and our conclusions are summarized in Section \ref{sec:conclusions}.

\section{Pair production in an electric field}

A constant electric field in (1+1) dimensions, $F_{\mu\nu} =
E\epsilon_{\mu\nu}$ with $E={\rm const}$, is invariant under
Poincar\'e transformations.  Its energy-momentum tensor has the vacuum
form, $T_\mu^{\;\nu} = \frac{1}{2}E^2 \delta_\mu^{\;\nu}$.  If the
field is coupled to particles of electric charge $e$ and mass $m$,
particle-antiparticle pairs will be spontaneously produced with a
critical separation $d_c= 2m/e E$,  
%where $\bar E = E - e/2$, 
so that the rest energy of the particles, $2m$ is compensated by the potential energy, $-eE d_c$.  (For definiteness we assume that $E$ and $e$ are both positive.)  The electric field in the space between the pair is reduced to the value $(E-e)$, and the vacuum energy is reduced accordingly.  The created particles play the role of bubble walls in this model.
The rate of pair creation per unit length is (see, e.g., Ref.~\cite{Cohen} and references therein)\footnote{This expression is exact in external field (corresponding to the limit $e \ll E$) . Backreaction can easily be taken into account in the instanton approximation (see, e.g., \cite{Leblond:2009fq}). In this case, the result is given by (1), but with $E$ replaced by $\bar E= E-e/2$. The instanton approximation is valid in the limit $m^2 \gg eE$.}
\beq
\Gamma = \frac{e E}{2\pi}\exp\left(-\frac{\pi m^2}{e E}\right).
\label{2Drate}
\eeq

If the particles are coupled to themselves or to another particle species, then pairs separated by a distance $d>d_c$ can be produced.  Such pairs have negative energy, and their formation must be accompanied by production of compensating positive-energy particles.  It is clear that the boost integration argument is precisely concerned with this situation.  If the argument is correct, the pair production rate should diverge when particle interactions are included.  We shall now examine whether or not such a divergence actually occurs.

\section{Massless pairs}
\label{class}

Massless QED in (1+1) dimensions, also known as the Schwinger model, is described by the action
\begin{equation}
S=\int d^2 x \left[\bar\psi \gamma^\mu(i\partial_\mu+eA_\mu)\psi - {1\over 4} F_{\mu\nu}F^{\mu\nu}\right]. 
\label{action}
\end{equation}
It is equivalent to the bosonic theory \cite{CJS}
\begin{equation}
S=\int d^2 x \left[{1\over 2}(\partial\phi)^2+  g A_\mu \epsilon^{\mu\nu} \partial_\nu \phi - {1\over 4} F_{\mu\nu}F^{\mu\nu}\right],
\label{actionb}
\end{equation}
where $g=e/\sqrt{\pi}$.  

In terms of the boson, the electric current is given by
\begin{equation}
j^{\mu} = e \bar\psi \gamma^\mu \psi = g \epsilon^{\mu\nu} \partial_\mu\phi.
\label{j}
\end{equation}
The number density of charge carriers $n$ is related to the electric current $j = -g{\dot\phi}$ by $j= 2e\ n$. It follows that the pair production rate is given by 
\begin{equation}
\Gamma = {1\over 2e}{dj\over dt} = -{\ddot\phi\over 2\sqrt{\pi}}.  
\label{rate}
\end{equation}

The field equation for $\phi$ is
\begin{equation}
\Box \phi = -gE, \label{accel}
\end{equation}
where $E$ is the electric field.  For a constant external field, $\phi$ develops an expectation value 
\begin{equation}
\langle\phi\rangle = \phi_0(t) = -{1\over 2} g E t^2,
\label{phi0}
\end{equation} 
and Eq.~(\ref{rate}) gives 
\begin{equation}
\Gamma = {e\over 2\pi}E .
\label{2Dmassless}
\end{equation}
As one might expect, there is no exponential suppression of pair production for massless fermions.

We note that the massless pair production rate (\ref{2Dmassless}) is in agreement with the zero-mass limit of Eq.~(\ref{2Drate}).  (For a detailed discussion of this limit, see \cite{Cohen}.)  It was first derived by Witten \cite{Witten} in the context of superconducting cosmic strings.

In this paper we will not be interested in the effect of back-reaction of the created pairs on the electric field.  However, if needed, this effect can be easily taken into account \cite{Cohen,Tanmay}.  From Maxwell's equation
\beq
{\dot E} = -j = g{\dot\phi} ,
\end{equation}
and using (\ref{accel}) we have
\beq
E = E_0 + g\phi
\label{E}
\eeq
and
\begin{equation}
\Box\phi + g^2 \phi = -gE_0 ,
\label{wave}
\end{equation}
where $E_0 = const$.  The solution of this equation for a spatially homogeneous $\phi$ with the initial condition $\phi = {\dot\phi} = 0$ at $t=0$, is 
\beq
\phi_0(t) = -g^{-1} E_0 [1-\cos(gt)],
\label{backreaction}
\eeq
and Eq.~(\ref{E}) gives
\beq
E = E_0 \cos(gt).
\eeq
This shows that the screening of the electric field by the produced pairs occurs on a timescale $t \sim g^{-1}$.  At smaller times $t\ll g^{-1}$, Eq.~(\ref{backreaction}) is well approximated by Eq.~(\ref{phi0}).

The above results are valid to all orders in $e$, and show that catastrophic vacuum decay does not occur in the Schwinger model.  One may be concerned that this model is somewhat special. Note that there are no propagating photons in (1+1) dimensions. Hence, if we neglect the dynamics of the electric field, the model does not include any particles that could
compensate for the increased negative energy of the pairs.  On the other hand, if we do include a dynamical electric field, the model is still equivalent to a free bosonic theory, and hence it may be considered non-generic. In order to account for more generic situations with
propagating particles in final states, we now consider more complicated interacting theories.

\subsection{Interactions}

The Schwinger model (\ref{action}) can be modified by adding self-interaction terms for the field $\psi$.  We first consider a simple interaction of the form 
\beq
{\cal L}_I = \lambda\ j_\mu j^{\mu}, 
\label{Thirring}
\eeq
where $\lambda$ is a coupling constant.  Apart from the Maxwell field, the action (\ref{action}) with this interaction term corresponds to the massless Thirring model.
Naively, the four fermion interaction could lead to pair creation processes which are not boost invariant, so one might expect catastrophic vacuum decay once an electric field is applied. 

This, however, does not happen.  In the bosonized form of the model, the interaction term (\ref{Thirring}) becomes
\beq
{\cal L}_I = \lambda g^2 (\partial \phi)^2,
\eeq
and amounts to a finite renormalization of the kinetic term for $\phi$.  Thus the model is equivalent to the bosonic Schwinger model, with the replacement $g \to g'= (1+2\lambda g^2)^{-1/2}g$. If we apply a constant
electric field, the electric current will grow at a constant rate, as before:
\begin{equation}
{dj\over dt}= {e\over 2 \pi} {eE\over (1+2\lambda g^2)}.
\end{equation}
Clearly, no catastrophic vacuum decay occurs in this model.
 
Turning now to more generic interactions, we introduce an additional scalar field $\chi$, which is coupled to $\psi$ via
\beq
{\cal L}_I = H(\chi) \epsilon^{\mu\nu}\partial_\mu j_\nu,
\eeq
where the function $H(\chi)$ is assumed to be ${\cal O}(\chi^2)$ at small $\chi$ and is otherwise arbitrary.  This form of interaction is chosen so that the model can be rewritten in an equivalent bosonic form.  We could also choose a coupling $G(\chi) j_\mu j^\mu$, but since the current has an expectation value linearly growing with time (at least to the lowest order in the interaction), this would result in a time-dependent, growing mass for $\chi$. 

Using Eq.~(\ref{j}) to express the current in terms of the bosonic field $\phi$, we have
\beq
\epsilon^{\mu\nu}\partial_\mu j_\nu = 2g\Box\phi,
\eeq
so the bosonic action takes the form
\begin{equation}
S=\int d^2 x \left[ {1\over 2}[(\partial \phi)^2+(\partial \chi)^2]-V(\chi)+ H(\chi) \Box \phi -gE\phi\right] ,
\end{equation} 
where we have also added a self-interaction potential for $\chi$.

To lowest order in the interactions, the expectation values of $\phi$ and $\chi$ are
$\langle\phi\rangle =\phi_0(t)$ and $\langle\chi\rangle=0$, with $\phi_0(t)$ from Eq.~(\ref{phi0}). 
Introducing $\hat \phi\equiv\phi -\phi_0$, the action transforms into
\begin{equation}
S=\int d^2 x \left[ {1\over 2}[(\partial\hat \phi)^2+(\partial \chi)^2]-V(\chi)+ H(\chi) (\Box\hat \phi-gE)\right] .
\label{hat}
\end{equation}  
For a constant electric field, the term proportional to $gE$ amounts to a finite renormalization of the mass and self-couplings of $\chi$.  

We see that even though the classical solution $\phi_0(t)$ is growing with time, the model (\ref{hat}) describing the quantum theory on this background has a time-independent Lagrangian and shows no signs of instability.  Once again, it seems clear that nothing catastrophic can happen to the vacuum in this model.

\section{Massive QED}\label{sec:massive}

We finally consider pair production in massive QED, with photons playing the role of the compensating positive-energy particles.  Since there are no photons in 2D, we consider the 4-dimensional case.  A constant, spatially homogeneous electric field is invariant under longitudinal boosts, so according to the boost integration argument, the rate for processes like
\beq
{\rm Vacuum} \to e^+ e^- \gamma , ~~~~~ {\rm Vacuum} \to e^+ e^- \gamma\gamma,
\label{photons}
\eeq
etc., should include integration over such boosts and should therefore be divergent.  

The total vacuum decay rate, including all particle production processes, is related to the imaginary part of the vacuum energy density,
\beq
\Gamma_{\rm vac} = 2 {\rm Im} \rho_{\rm vac},
\label{Gammavac}
\eeq
and can be found by evaluating the diagrams shown in Fig.~1.  Solid lines in these diagrams stand for electron propagators in a constant external field, and wavy lines represent photon exchange.  The one-loop diagram in Fig.~1(a), which does not include any virtual photon lines, gives the Heisenberg - Euler - Schwinger (HES) result,
\beq
\Gamma^{(1)} = \frac{(eE)^2}{4\pi^3}\exp\left(-\frac{\pi m^2}{eE}\right).
\label{HES}
\eeq
The higher-loop diagrams of Figs.~1(b)-(c) account for the processes
(\ref{photons}), where pair production is accompanied by photons, as
well as for the radiative corrections to the basic pair production
process, ${\rm Vacuum} \to e^+ e^-$.  The diagram in Fig.~1(d)
represents correlated formation of two pairs and radiative corrections
to the photon propagator.  If the boost integration argument is correct, the
multi-loop diagrams should diverge.  

\begin{figure}
  \begin{center}
 \includegraphics[width= 15 cm]{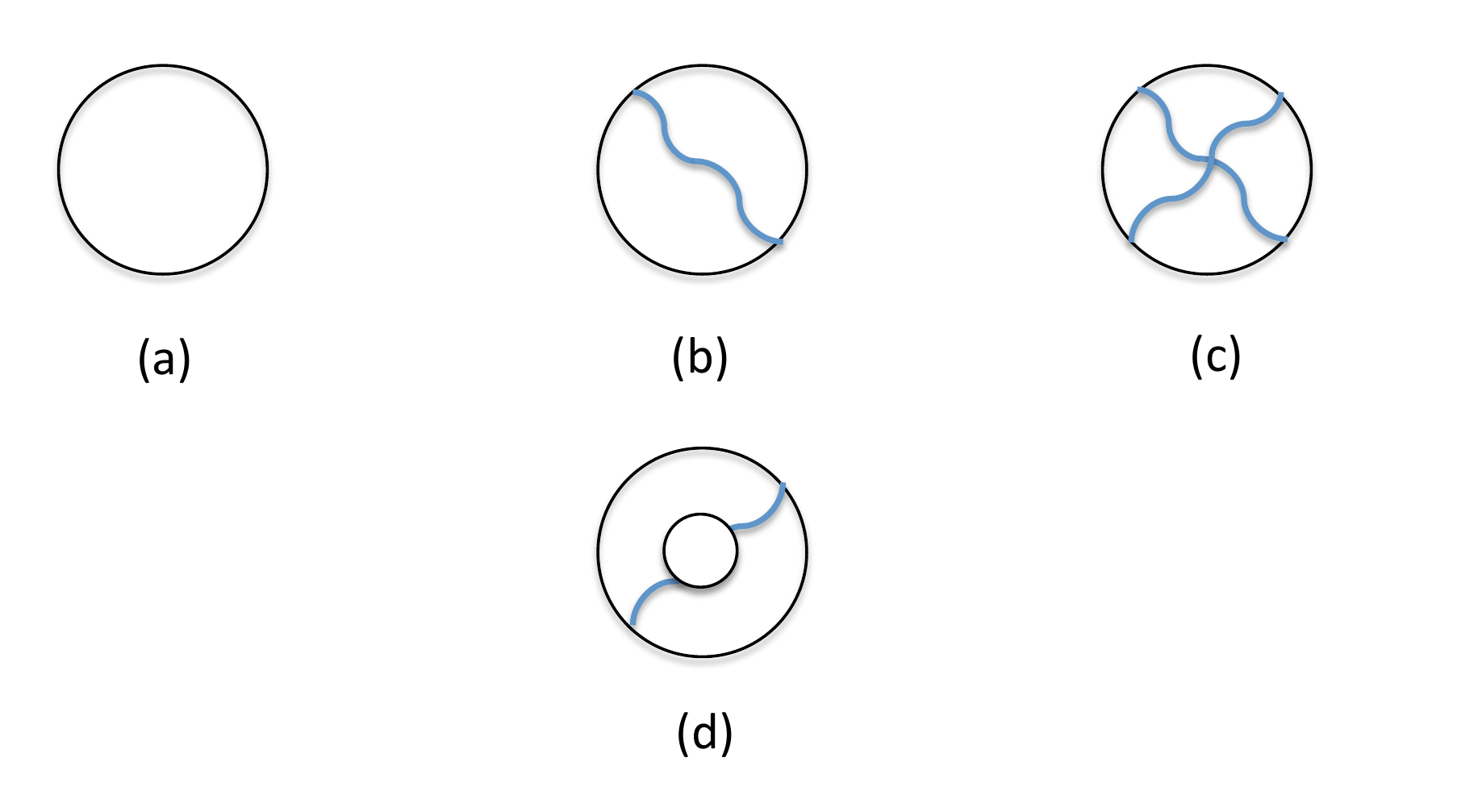}
  \end{center}
  \caption{Diagrams contributing to the decay of an electric field. The solid lines represent the electron propagator in the presence of the electric field, and the wavy lines
  represent the photon propagator.}
   \label{MAdSl}
\end{figure}

It should be noted that the pair production rate $\Gamma$ is generally different from the vacuum decay rate $\Gamma_{\rm vac}$ defined by Eq.~(\ref{Gammavac}).  At the one-loop level, the exact result for $\Gamma$ is given by Eq.~(\ref{HES}), while $\Gamma_{\rm vac}$ is \cite{Schwinger}
\beq
\Gamma_{\rm vac} = \frac{(eE)^2}{4\pi^3} \sum_{n=1}^\infty \frac{1}{n^2} \exp\left(-\frac{n\pi m^2}{eE }\right).
\label{S}
\eeq
For an illuminating discussion of this distinction, see, e.g., Ref.~\cite{Cohen}.\footnote{The distinction is particularly important in the massless (1+1)-dimensional case, when 
$\Gamma_{\rm vac}$ becomes formally divergent, while the pair production rate (\ref{2Dmassless}) remains finite \cite{Cohen}.}  Here, we will be interested in the weak field limit,
\beq
E\ll m^2/e,
\label{weakfield} 
\eeq
when $\Gamma\approx\Gamma_{\rm vac}$ to very good accuracy.  Hence, we will not distinguish between $\Gamma_{\rm vac}$ and $\Gamma$ in what follows.

Higher-loop corrections to the HES formula (\ref{HES}) have been extensively studied in the literature.  Affleck, Alvarez and Manton (AAM) \cite{AAM} derived a remarkably simple formula, 
which includes contributions of all multi-loop diagrams containing a single electron loop,
\beq
\Gamma^{(all~loop)} = \frac{(eE)^2}{4\pi^3}\exp\left(-\frac{\pi m^2}{eE} +\frac{e^2}{4}\right).
\label{AAM}
\eeq
It was argued by AAM that diagrams with more than one electron loop are subdominant in the weak field limit (\ref{weakfield}); then Eq.~(\ref{AAM}) is the full multi-loop pair creation rate in this limit.
Comparing it to (\ref{HES}), we see that the correction term is definitely finite and small, as one might naively expect. 

The AAM derivation was not rigorous, as it relied on the steepest descent evaluation of the path integral for the QED effective action and on a somewhat cavalier treatment of mass renormalization.  However, direct calculations \cite{LebedevRitus84,Ritus,DunneSchubert00} of the two-loop contribution in Fig.~1(b) also yielded a finite result, which is in agreement with (\ref{AAM}).

It may be instructive to consider how the infinite boost integration plays out in the QED context.  In the semiclassical approximation, electrons (and positrons) move along hyperbolic trajectories with a constant proper acceleration, $a=eE/m$.  The characteristic energy of the emitted photons in the rest frame of the electron is $\omega\sim a$, and in the weak field limit (\ref{weakfield}) we have $\omega\ll m$.  This indicates that photon emission has negligible effect on the motion of the electrons, and one can simply consider radiation from a charged particle moving along a fixed classical trajectory.

For a charge in hyperbolic motion, photons will be emitted with the same characteristic energy in the rest frame of the charge, but will be boosted to arbitrarily high energies in the observer's frame.  However, the transverse momentum of the photons, in the plane perpendicular to the electric field, will not be boosted.  Hence, the appropriate quantity to calculate is the probability of emitting a photon with a given transverse momentum ${\bf p}_\perp$.  This calculation has been done, in the lowest order of perturbation theory, in Ref.~\cite{Higuchi} for photons and in Ref.~\cite{RenWeinberg} for the emission of massless spin-zero particles by a scalar source.  The resulting probability is proportional to the divergent integral over the particle's proper time (which can also be thought of as an integral over boosts, since a boost causes a shift in the proper time of emission).  How do we interpret this divergence?

Of course, the probability of photon emission cannot be greater than one, so the divergence indicates a breakdown of the perturbation theory.  The origin of the problem can be understood if we consider a situation where the electric field is turned on only for a finite period of time $T$. We shall assume this period to be short enough, so that the probability of emitting more than one photon in time $T$ is small.  (Note that we can treat the electric charge $e$ as a free parameter.  For small values of $e$, photons will be emitted very rarely, so the time $T$ can be very long.)  In this case there will be no divergence.  The probability of photon emission will be proportional to the proper time $\tau$ spent by the charge in the electric field.  This simply reflects the fact that there is a fixed photon emission probability per unit proper time. 

What happens when we increase the time $T$?  For very large values of $T$, the probability of emitting a single photon becomes very small.  Photon emission events along the particle's trajectory can be regarded as independent, so the probability for emitting $n$ photons will be given by the Poisson distribution
\beq
P_n = \frac{{\bar n}^n}{n!} e^{-{\bar n}},
\eeq
where ${\bar n}$ is the average number of the emitted photons (which is proportional to $\tau$).  In the limit $T\to\infty$, we have ${\bar n}\to\infty$, so the probability of emitting any finite number of photons goes to zero.  On the other hand, ${\bar n}$ is proportional to $e^2$, so a formal perturbative expansion gives a divergent result for the emission of a single quantum, $P_1^{(pert)} = {\bar n} + {\cal O}(e^4)$.\footnote{The situation here is similar to the divergence encountered in emission of soft photons, known as the infrared catastrophe in QED; see, e.g., \cite{AkhiezerBerestetsky}.}

\section{Conclusions}\label{sec:conclusions}

In this paper we have analyzed Dvali's boost integration argument indicating that a Minkowski vacuum can either be absolutely stable or cannot exist at all.  Any decay channel allowing negative-energy bubbles would supposedly result in a divergent vacuum decay rate.  This would imply good news for inhabitants of a Minkowski vacuum: if your vacuum survived for any finite time, it is guaranteed to endure forever. 
Since our vacuum is close to Minkowski, by continuity we should then expect it to be extremely long lived.

Here, we argued that if the boost integration argument is correct, it should also apply to pair production of charged particles in a constant electric field.  The pair production rate should then become infinite when processes like (\ref{photons}), where the pairs are produced together with some other particles, are included.  

We do not know of any calculations that would exhibit such a divergence.  On the other hand, here we have presented evidence that such divergences do not occur in a class of massless (1+1)-dimensional theories and in (3+1)-dimensional (massive) QED.

The divergence found in the boost integration argument arises due to the integration over the
momenta of the vacuum decay products. A perturbative calculation of the
probability for nucleating a negative energy bubble accompanied by a single
positive energy blob of matter in the final state contains a divergent
integral over the momentum of the blob.

However, as we discussed in Section IV, this divergence only indicates a
breakdown of perturbation theory. Physically, the reason is the following.
Fluctuations such as the emission of a blob by an expanding bubble occur
with a constant probability per unit wall area per unit proper time. The
boost integral in the naive perturbative calculation only changes the value
of proper time on the worldsheet at which a blob is emitted. If the bubble
expands for a finite time, i.e., short enough that the emission of one blob is
unlikely, then perturbation theory applies, and the boost integral will
correctly account for the fact that the probability of emission is linear in
the proper time interval available. However, it is clear that this integral
is unrelated to the probability of nucleating the bubble itself. As the
bubble expands, the probability of having the bubble with the blob
increases, and the probability of having the bubble without the blob
decreases. Once the probability for having more than one blob is
significant, perturbation theory has broken down.

In a Lorentz invariant situation, where the bubble expands for an infinite
amount of time, the correct answer for the probability of having just one
blob (or any finite number of them) in the final state is actually zero. In
other words, the final state consists of a Lorentz
invariant distribution of infinitely many blobs, generated as the bubble
expands into the false vacuum. But of course this does not imply an infinite
rate of bubble nucleation.

A metastable Minkowski vacuum cannot exist forever and thus cannot be truly
Poincar\'e invariant.  However, such vacua can be formed as bubbles in the
inflationary multiverse and will decay by producing AdS
bubbles.  Our analysis indicates that the corresponding decay rate can be
calculated using the standard Coleman-de Luccia method and does not involve
any divergent boost integrations.\footnote{The bubble nucleation rate can be affected by the initial conditions,
representing how the metastable vacuum was set up.  However, this is a
transient effect, which we expect to be important on a timescale of the
order of the instanton size.  Note that in the massless limit of QED, when
the size of the instanton is zero, the pair creation rate depends only on
the local value of the electric field, regardless of the initial
conditions.}

\subsection*{Acknowledgements}
We are grateful for comments by and discussions with Gia Dvali, Gregory Gabadadze, Slava Mukhanov, and Tanmay Vachaspati.
This work was supported in part by grants DURSI 2009 SGR 168, MEC FPA 2007-66665-C02 and CPAN CSD2007-00042 Consolider-Ingenio 2010 (JG) and by the National Science Foundation grant PHY-0855447 (BS and AV).

\end{document}